\documentclass[%
 aip,
 amsmath,amssymb,
 reprint,%
]{revtex4-1}

\usepackage{graphicx}
\usepackage{dcolumn}
\usepackage{bm}

\usepackage[utf8]{inputenc}
\usepackage[T1]{fontenc}
\usepackage{mathptmx}
\usepackage{etoolbox}

\makeatletter
\def\@email#1#2{%
 \endgroup
 \patchcmd{\titleblock@produce}
  {\frontmatter@RRAPformat}
  {\frontmatter@RRAPformat{\produce@RRAP{*#1\href{mailto:#2}{#2}}}\frontmatter@RRAPformat}
  {}{}
}%
\makeatother
\begin{document}

\preprint{AIP/123-QED}

\title[]{Observation and formation mechanism of 360$^{\circ}$ domain wall rings in Synthetic Anti-Ferromagnets with interlayer chiral interactions}
\author{Miguel A. Cascales Sandoval}
\affiliation{SUPA, School of Physics and Astronomy, University of Glasgow, Glasgow G12 8QQ, UK}%

\author{A. Hierro-Rodr{\'i}guez*}%
\email[Corresponding author e-mails: ]{hierroaurelio@uniovi.es, amaliofp@unizar.es.}
\affiliation{Departamento de F{\'i}sica, Universidad de Oviedo, 33007, Oviedo, Spain}
\affiliation{CINN (CSIC-Universidad de Oviedo), 33940, El Entrego, Spain}

\author{S. Ruiz-G{\'o}mez}
\affiliation{Max Planck Institute for Chemical Physics of Solids, 01187 Dresden, Germany}

\author{L. Skoric}
\affiliation{University of Cambridge, Cambridge CB3 0HE, UK}

\author{C. Donnelly}
\affiliation{Max Planck Institute for Chemical Physics of Solids, 01187 Dresden, Germany}

\author{M. A. Niño}
\affiliation{ALBA Synchrotron Light Facility, 08290
Cerdanyola del Valles, Spain}

\author{Elena Y. Vedmedenko}
\affiliation{Institute of Applied Physics, University of Hamburg, Hamburg, Germany}

\author{D. McGrouther}
\affiliation{SUPA, School of Physics and Astronomy, University of Glasgow, Glasgow G12 8QQ, UK}

\author{S. McVitie}
\affiliation{SUPA, School of Physics and Astronomy, University of Glasgow, Glasgow G12 8QQ, UK}

\author{S. Flewett}
\affiliation{Instituto de F{\'i}sica, Pontificia Universidad Cat{\'o}lica de Valpara{\'i}so, Avenida Universidad 330, Valpara{\'i}so, Chile}

\author{N. Jaouen}
\affiliation{SOLEIL Synchrotron, L'ormes des
Merisiers, 91192 Gif-Sur-Yvette, Cedex, France}

\author{M. Foerster}
\affiliation{ALBA Synchrotron Light Facility, 08290
Cerdanyola del Valles, Spain}

\author{A. Fern{\'a}ndez-Pacheco*}
\affiliation{SUPA, School of Physics and Astronomy, University of Glasgow, Glasgow G12 8QQ, UK}
\affiliation{Instituto de Nanociencia y Materiales de Arag{\'o}n, CSIC-Universidad de Zaragoza, 50009 Zaragoza, Spain}


\begin{abstract}

The Interlayer Dzyaloshinskii-Moriya interaction (IL-DMI) chirally couples spins in different ferromagnetic layers of multilayer heterostructures. So far, samples with IL-DMI have been investigated utilizing magnetometry and magnetotransport techniques, where the interaction manifests as a tunable chiral exchange bias field. Here, we investigate the nanoscale configuration of the magnetization vector in a synthetic anti-ferromagnet (SAF) with IL-DMI, after applying demagnetizing field sequences. We add different global magnetic field offsets to the demagnetizing sequence in order to investigate the states that form when the IL-DMI exchange bias field is fully or partially compensated. For magnetic imaging and vector reconstruction of the remanent magnetic states we utilize X-ray magnetic circular dichroism photoemission electron microscopy, evidencing the formation of 360$^{\circ}$ domain wall rings of typically 0.5-3.0 $\mu m$ in diameter. These spin textures are only observed when the exchange bias field due to the IL-DMI is not perfectly compensated by the magnetic field offset. From a combination of micromagnetic simulations, magnetic charge distribution and topology arguments, we conclude that a non-zero remanent effective field with components both parallel and perpendicular to the anisotropy axis of the SAF is necessary to observe the rings. This work shows how the exchange bias field due to IL-DMI can lead to complex metastable spin states during reversal, important for the development of novel spintronic devices.

\end{abstract}

\maketitle

The interlayer Dzyaloshinskii-Moriya interaction (IL-DMI) is an interlayer-mediated antisymmetric exchange interaction observed in multilayer heterostructures, which promotes orthogonal coupling between spins in different magnetic layers \cite{fernandez2019symmetry,han2019long,vedmedenko2019interlayer,avci2021chiral}. This coupling mechanism may find interesting applications in 3D nanomagnetism, as it provides the opportunity of inducing chiral magnetic states in a layer controlled by the magnetic state of another.

IL-DMI can be understood as an effective unidirectional magnetic field breaking the symmetry of the reversal process, leading to a chiral exchange bias \cite{han2019long}. This exchange bias field has been for instance utilized to achieve field free spin-orbit torque mediated deterministic switching of perpendicularly magnetized thin film systems \cite{huang2022growth,lau2016spin,he2022field,wang2022field,bekele2020enhanced}, as an alternative to exchange bias generated at ferromagnet/anti-ferromagnet interfaces. So far, samples with IL-DMI have been typically investigated with magnetometry and magnetotransport techniques such as the magneto-optical Kerr effect (MOKE) \cite{fernandez2019symmetry} and anomalous Hall-effect \cite{han2019long,wang2021spin,kammerbauer2022controlling}.

Here, the remanent magnetic domain configurations present in a synthetic anti-ferromagnet (SAF) with chiral exchange coupling due to IL-DMI are investigated using X-ray magnetic circular dichroism photo-emission electron microscopy (XMCD-PEEM). The states imaged are obtained after performing a demagnetizing \cite{jang2012formation,dean2011formation} process with different external field offsets added to the cycling sequence. By combining multiple XMCD-PEEM projections measured at different azimuthal angles, the magnetization vector is reconstructed, evidencing the formation of 360$^{\circ}$ domain wall (DW) rings. The IL-DMI is found to be key for the stability of these structures, as rings are only observed when the exchange bias IL-DMI field is not fully compensated by the external field offset. Experiments are complemented with micromagnetic simulations, which highlight the importance of the relative orientation of external and demagnetizing dipolar magnetic fields with the in plane anisotropy axis of the sample for the formation of the 360$^{\circ}$ DW rings.

The SAF structure under investigation consists of Si/Ta (4 nm)/Pt (10 nm)/Co (1 nm)/Pt (0.5 nm)/Ru (1 nm)/Pt (0.5 nm)/CoFeB (2 nm)/Pt (2 nm)/Ta (4 nm). The Co layer has strong perpendicular magnetic anisotropy, \textit{i.e.}, it is hard out-of plane (OOP), whereas the CoFeB (Co: 60$\%$, Fe: 20$\%$, B: 20$\%$) has been tailored to show moderately low in-plane (IP) anisotropy by tuning its thickness above its spin reorientation transition, \textit{i.e.}, it is soft IP. In this type of SAFs, a chiral exchange bias due to IL-DMI has been previously observed under measurement of minor IP hysteresis loops on which solely the CoFeB layer reverses while the Co stays fixed in the OOP direction \cite{fernandez2019symmetry}. Upon reversal of the orientation of the Co layer, the exchange bias changes sign. The unidirectional nature of the effect is manifested by the IP angular dependence of the bias, which shows opposite sign for the two possible directions parallel to the IP anisotropy easy axis (EA), and zero for the directions orthogonal to it. 

\begin{figure}[ht]
    \centering
    \includegraphics[scale=0.185]{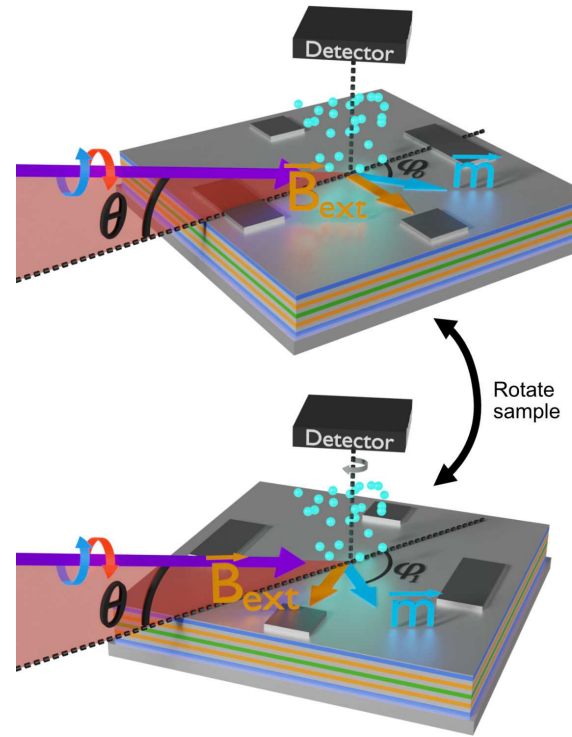}
    \caption{\label{fig:fig1} Schematic of two different relative sample/X-ray beam orientations. $\varphi_{0}$ and $\varphi_{1}$ are the different relative angles between the X-ray beam (purple arrow) and the magnetization vector (blue arrow and $\vec{m}$ symbol), $\theta$ is the incidence angle with respect to the surface plane of the X-rays, the blue and red circular arrows denote both X-ray circular polarization eigenmodes and $B_{ext}$ the external magnetic field direction provided by the dipolar electromagnet.}
\end{figure}

To investigate the effect of IL-DMI on the domain states forming in these samples, magnetic microscopy measurements were taken at CIRCE \cite{aballe2015alba} beamline in ALBA Synchrotron, using XMCD-PEEM (see sketch in figure \ref{fig:fig1}). In this setup, the azimuthal angle of the sample with respect to the X-rays can be modified in the full 360$^{\circ}$ angular range, while the polar angle is fixed to 16$^{\circ}$ with respect to the surface plane, giving large sensitivity to IP components. Both circular polarization images are recorded and combined exploiting XMCD for magnetic contrast \cite{stohr2006magnetism,beaurepaire2001magnetism,yokoyama2008magnetic}, measured at both Fe and Co's $L_{3}$ absorption edges at each magnetic state. Prior to the experiments, 50 nm thick $\text{Pt}_{x}\text{C}_{1-x}$ squares and rectangles were deposited using focused electron beam induced deposition (FEBID) on top of the film, serving as non-magnetic references for equalizing both circular polarization images in order to properly compute the final XMCD image. The holder used for mounting the sample into the PEEM chamber has an embedded dipolar electromagnet \cite{foerster2016custom,aballe2015alba}, giving the possibility of applying in-plane uniaxial magnetic fields ($\vec{B}_{ext}$). The nominal IP EA of the CoFeB ($\hat{x}$), given by the rectangle alignment mark's long axis, is mounted parallel to the electromagnet's axis after previously saturating the Co layer in one of the OOP directions.

\begin{figure*}
\centering
\includegraphics[scale=0.13]{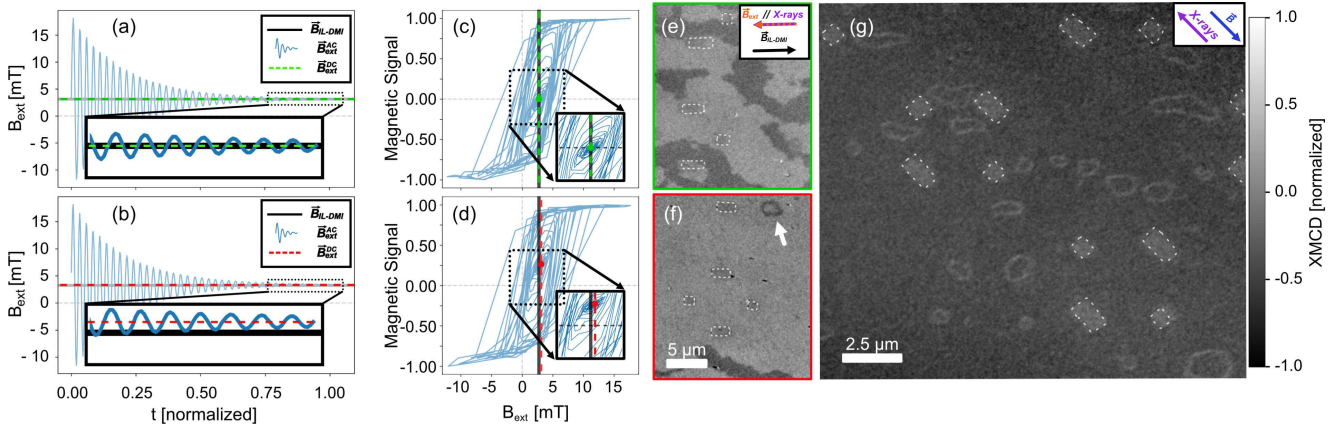}
\caption{\label{fig:fig2} (a,b) External field demagnetizing protocol, compensating either fully (a), or partially (b) the IL-DMI exchange bias field via a field offset ($B_{x} = 0$ and $B_{x} \neq 0$). (c,d) XRMS measurements taken with circular incident polarization during application of the demagnetization procedure, respectively for the field sequences in (a,b). (e,f) XMCD-PEEM images measured at final green and red cases at Co's $L_{3}$ edge. The inset in (e) is common to (f), representing the external magnetic field, X-ray and IL-DMI field directions. (g) XMCD-PEEM image taken at Co's $L_{3}$ edge after partially demagnetizing the CoFeB film and utilizing a larger FOV, evidencing the formation of a large number of rings. The inset here shows the direction of the X-rays and the net effective field acting on the sample. The white dashed lines in (e,f,g) denote the non-magnetic reference $\text{Pt}_{x}\text{C}_{1-x}$ squares and rectangles}
\end{figure*}

In order to obtain a magnetically non-trivial remanent state, a demagnetization protocol which consists of consecutively decreasing IP sinusoidal $\vec{B}_{ext}$ signals is applied. $\vec{B}_{ext}$ thus consists of an AC oscillating component ($\vec{B}^{AC}_{ext}$), and a DC or external field offset component ($\vec{B}^{DC}_{ext}$). The protocol is demonstrated in figure \ref{fig:fig2}, where the magnetization is probed in a separate X-ray resonant magnetic scattering (XRMS) experiment, performed at SEXTANTS beamline in SOLEIL synchrotron. For this, XRMS hysteresis loops are taken where the specularly reflected signal is recorded, using field sequences where the IL-DMI exchange bias field ($\vec{B}_{IL-DMI}$) is either fully or partially compensated by $\vec{B}^{DC}_{ext}$. The X-ray beam spot used is 300 $\mu$m in diameter \cite{jaouen2004apparatus}, and the X-ray beam was set to 13$^{\circ}$ of incidence with respect to the sample plane, giving large sensitivity to the IP magnetization \cite{jal2013reflectivite}.

Figures \ref{fig:fig2} (c,d) show the XRMS hysteresis loops for the two demagnetizing sequences in figures \ref{fig:fig2} (a,b). In the green (fully demagnetized case) shown in figure \ref{fig:fig2} (c), $\vec{B}_{IL-DMI}$'s $\hat{x}$ component is fully compensated by $\vec{B}_{ext}$, \textit{i.e.}, $B_{x} = 0$, with $\vec{B} = \vec{B}_{ext} - \vec{B}_{IL-DMI}$, yielding final zero net magnetic signal. This state corresponds to the one observed via XMCD-PEEM in figure \ref{fig:fig2} (e), where the area of black and white domains within the field of view (FOV) is indeed approximately equal. The situation is different for the red (partially demagnetized case) shown in figure \ref{fig:fig2} (d), where $\vec{B}_{IL-DMI}$'s $\hat{x}$ component is not fully compensated ($B_{x}\neq 0$), reaching a final state with non-zero net magnetic signal. This agrees with the corresponding PEEM experiment shown in figure \ref{fig:fig2} (f), where there is a clear dominating magnetic configuration aligned with $\vec{B}_{ext}$. In the partially demagnetized case, a number of ring-like magnetic textures (0.5-3.0 $\mu m$ in diameter) is frequently observed by PEEM, for instance, the one marked by the white arrow in figure \ref{fig:fig2} (f). The emergence of a large number of rings is found in partially demagnetized states as the one in figure \ref{fig:fig2} (g), where an area with a larger field of view is shown. All the images shown here are taken at the Co $L_{3}$ edge and refer to states forming on the CoFeB layer, since the magnetic features are identical at both Fe and Co edges, with Co showing a significantly better signal-to-noise ratio due to the stoichiometry of the CoFeB layer (see supplementary material).

\begin{figure}[ht]
    \centering
    \includegraphics[scale=0.19]{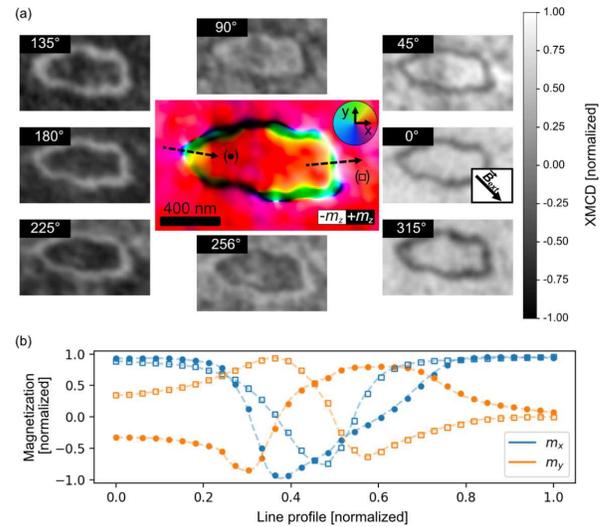}
    \caption{\label{fig:fig3} (a) XMCD-PEEM based vector reconstruction of the magnetization of a 360$^{\circ}$ DW ring using 8 projections. Gray-scale images correspond to the 8 XMCD-PEEM projections taken at different azimuthal X-ray beam/sample relative angles. The angle of each projection is given with respect to the x-axis indicated in the central figure. The applied external magnetic field direction is given by the inset arrow of the 0$^{\circ}$ projection. The central color-map represents the spatially resolved magnetic vector configuration, utilizing the hsl colormap for the IP directions, and black and white for the OOP directions. (b) IP components of $\vec{m}$ along the profiles given by the dashed arrows overlayed on the central color plot, evidencing 360$^{\circ}$ DW.}
\end{figure}

To understand better the magnetic configuration of these spin textures, vector imaging of one of the rings is performed by measuring several X-ray beam/sample projections. For this, the sample is rotated in the PEEM chamber with respect to the  X-ray direction, as shown in figure \ref{fig:fig1}. XMCD images are obtained for a total of 8 azimuthal angles, see figure \ref{fig:fig3} (a). The images have been previously rotated and aligned with respect to each other in order to have the same spatial orientation, following a similar procedure to the one described in \cite{le2012studying}. Additionally, deformations in different projection images are corrected by an algorithm which makes use of a combination of image registration techniques, using the $\text{Pt}_{x}\text{C}_{1-x}$ marker's geometrical shape as reference landmarks. These deformations arise from the fact that the used PEEM microscope does not inherently present circular symmetry. Some of the electron optical elements, in particular the 120$^{\circ}$ beam splitter, typically introduce image distortions which become relevant when overlaying images for different sample orientations.

The spatially resolved normalized magnetization vector ($\vec{m}$) is then reconstructed \cite{le2012studying,ruiz2018geometrically,ghidini2022xpeem,scholl2002x,chopdekar2013strain,chmiel2018observation,digernes2020direct}, by performing a pixel-by-pixel least squares fitting of the XMCD profile (given by $\vec{k}\cdot\vec{m}$, with $\vec{k}$ being the incident X-ray wave-vector) as a function of the azimuthal rotation angle. The resulting vector directions of $\vec{m}$ are shown in the central image of figure \ref{fig:fig3} (a). The line profiles in figure \ref{fig:fig3} (b) evidence the presence of 360$^{\circ}$ DWs separating the outer and inner domains, whose orientation is the same, and the direction of $\vec{m}$ within them is the result of the EA and $\vec{B}$ directions. In the reconstruction, great precision is achieved in the determination of the IP components due to the grazing incidence of the X-rays, which on the other hand reduces the sensitivity to OOP components. Furthermore, the magnetic signal decreases in areas where the magnetization changes over small spatial lengthscales in comparison with the microscope's resolution, giving a larger uncertainty in the domain walls' OOP components.

360$^{\circ}$ DW rings have been previously observed in individual ferromagnetic Permalloy thin film layers \cite{smith1962noncoherent}, in multilayered heterostructures \cite{heyderman1991360,heyderman1994tem,portier2000formation}, in exchanged biased films \cite{schafer1993domain,cho1999characteristics,dean2011formation} and in magnetic tunnel junctions \cite{gillies1995magnetization}. The role of Bloch lines \cite{heyderman1991360}, DW splitting \cite{heyderman1994tem} and dispersion in anisotropy  \cite{dean2011formation} have been proposed as mechanisms for explaining their formation. 

\begin{figure*}
\centering
\includegraphics[scale=0.136]{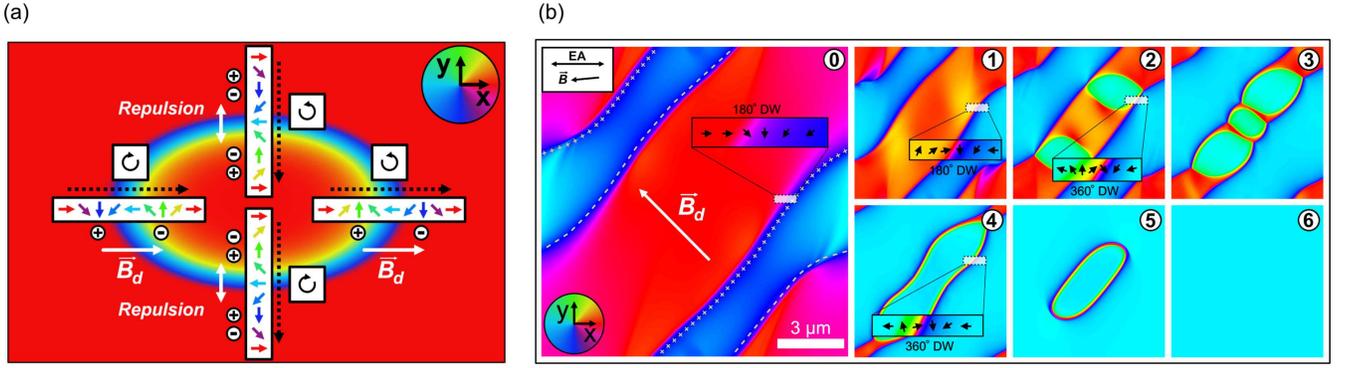}
\caption{\label{fig:fig4} (a) Simplified sketch of a 360$^{\circ}$ DW ring as the one obtained in the reconstruction shown in figure \ref{fig:fig3}, where the magnetization winding sense at 4 different locations is illustrated by the colored arrows. The rotating arrows denote the sense of rotation of magnetic spins along the dashed line arrows. (b) Dynamical evolution (numbered states) of the simulation space to study the formation of 360$^{\circ}$ DW rings with IP magnetization vector directions given by the hsl colormap. EA and magnetic field $\vec{B}$ directions are indicated by the inset arrows in state 0. In state (0), $\vec{B}_{d}$ represents the demagnetizing field arising from the magnetic charges "+" and "-", which is non-homogeneous and becomes stronger as the domain walls get closer to each other (subsequent states). The time for each state is given relatively to state (0); $t_{1} = 0.10$ ns, $t_{2} = 0.22$ ns, $t_{3} = 0.42$ ns, $t_{4} = 0.66$ ns, $t_{5} = 1.90$ ns and $t_{6} = 3.40$ ns.}
\end{figure*}

In order to determine the dependence between the stability of 360$^{\circ}$ DW rings and the IL-DMI effective field, focus is first set on their topology and magnetic charge distribution. For this, the simplified sketch of the vector reconstructed configuration shown in figure \ref{fig:fig4} (a) is used, assuming a fully IP texture. The figure shows an elongated, 360$^{\circ}$ DW ring with the EA along $\vec{x}$. A 1D 360$^{\circ}$ DW is a topologically non-trivial structure, having been sometimes denominated 1D skyrmions in the literature \cite{cheng2019magnetic}. However, when these DW form a ring structure as the one shown here, the net winding number computed along a line profile diametrically crossing the whole ring is zero due to the opposite chirality of the two encountered 360$^{\circ}$ DW. Therefore, a 360$^{\circ}$ DW ring as a whole is topologically trivial, which means that it can be continuously deformed into a single domain state.

Additionally, the elongation observed for the ring textures along the EA is expected based on magnetostatics arguments, \textit{i.e.}, Néel walls prefer to align parallel to the EA rather than perpendicular \cite{hubert2008magnetic} to it. Furthermore, the magnetic charge distribution along the ring structure is anisotropic, as evidenced by the "+" and "-" in the figure. For both (top and bottom) walls parallel to the EA, the two 180$^{\circ}$ DW forming the 360$^{\circ}$ DW repel each other, which in combination with the ferromagnetic exchange promote the growth of the annular domain (cyan). In the absence of a net magnetic field, this magnetic texture would expand and relax into broader domains separated by consecutive 180$^{\circ}$ DW \cite{muratov2008theory}. This contrasts with the (left and right) DW orthogonal to the anisotropy axis, where the demagnetizing field arising from the charge distribution now makes the two consecutive 180$^{\circ}$ DW attract each other, opposing the expansion of the annular domain promoted by ferromagnetic exchange. The competition of these interactions dictates the stability of the vertical 360$^{\circ}$ DW component at zero field \cite{muratov2008theory}. Thus, for the overall 360$^{\circ}$ DW ring texture to remain stable, a net non-zero $\vec{B}$ field along the easy axis is required to compress from both inner and outer parts, preventing it from relaxing and expanding into broader domains, \textit{i.e.}, two consecutive domains separated by 180$^{\circ}$ DWs. Additionally, this field prevents the unwinding of the inner domain due to it being topologically trivial. This agrees well with previous observations of 360$^{\circ}$ DW rings in systems with exchange bias \cite{schafer1993domain,cho1999characteristics,dean2011formation}, since the field sequences used there do not add field offsets, resulting in net non-zero fields at remanence.

Finally, we focus on investigating the mechanism behind the nucleation of the rings using micromagnetic simulations. For this, solely a single IP thin film is modeled representing the CoFeB layer of the SAF, whose simulation parameters are summarized in table \ref{tab:table1}. As in experiments, $\vec{B}_{IL-DMI}$ is parallel to the EA (horizontal $\vec{x}$ direction), and it is modeled through the effective field $\vec{B}$.

\begin{table}[ht]
\caption{\label{tab:table1}%
Parameters for the micromagnetic simulation. Values for parameters were respectively found at: $\alpha$\cite{glowinski2017cofeb}, $M_{S}$\cite{fernandez2019symmetry}, $A_{exc}$\cite{devolder2016exchange} and $K$\cite{fernandez2019symmetry}.}
\begin{ruledtabular}
\begin{tabular}{cc}
Parameter & Value\\
\colrule
Cells (x,y,z) & 3072 $\times$ 3072 $\times$ 1\\
Cell size (x,y,z) & 4.5 $\times$ 4.5 $\times$ 5.0 [nm]\\
$\alpha$ & 0.02\\ 
$M_{S}$ & 1.2$\cdot 10^{6}$ [A/m]\\ 
$A_{exc}$ & 20 [pJ/m]\\ 
$K$ & $1.8\cdot 10^{3}$ [J/$m^{3}]$\\ 
Defect size &  0.8 [$\mu$m]\\
Defect anisotropy & $3.6\cdot 10^{2}$ [J/$m^{3}]$\\
Periodic boundary conditions & $16\times 16$ [repetitions]\\
\end{tabular}
\end{ruledtabular}
\end{table}

In order to model the formation process via micromagnetic simulations, the first step consists of obtaining a magnetic configuration formed by multiple domains, resembling an intermediate state during the demagnetizing protocol. For this, the simulation starts from a fully saturated $+m_{x}$ configuration. Then, a $\vec{B}$ field of increasing magnitude is applied along $-\hat{x}$ with a small $-\hat{y}$ component (5$^{\circ}$ with respect to $+\hat{x}$), which represents a misalignment between $\vec{B}_{ext}$ and the EA. A circular defect of 0.8 $\mu$m radius and with 20 $\%$ anisotropy value of the layer is included to trigger the nucleation of domains (cyan). Once $\vec{B}$ reaches the switching field magnitude, the system is allowed to evolve dynamically, eventually reaching the state shown in [figure \ref{fig:fig4} (b), state (0)]. All 180$^{\circ}$ DWs in the simulation space have a $-m_{y}$ component set by $B_{y}$ \cite{jang2012formation}, resulting in the magnetic charge distribution, once again represented by the "+" and "-" signs.


As DW of opposite charge get closer promoted by the growth of the inverted domain, the intensity of the demagnetizing field ($\vec{B}_{d}$) associated to the magnetic charges begins to increase [figure \ref{fig:fig4} (b), state (1)]. The varying $\vec{B}_{d}$ in combination with the constant $\vec{B}$, eventually reaches the sufficient field magnitude for overcoming the anisotropy dominated nucleation field, as a consequence locally inverting the magnetization. This allows for the formation of a new pair of 180$^{\circ}$ DWs during the reversal, with their core magnetization pointing along $+m_{y}$, given that the $\hat{y}$ component of $\vec{B}_{d}$ in this case is opposite and larger in magnitude than the original set by $\vec{B}$, \textit{i.e.}, $B_{d_{y}}$ > - $B_{y}$. This is exemplified in the crop shown in [figure \ref{fig:fig4} (b), state (2)]; the new 180$^{\circ}$ DW forms next to the original $\rightarrow\downarrow\leftarrow$ from [figure \ref{fig:fig4} (b), state (1)], resulting in a 360$^{\circ}$ DW with $\leftarrow\uparrow\rightarrow\downarrow\leftarrow$ configuration. Thus, for the formation of 360$^{\circ}$ DW and the forthcoming rings, it is key for the $\hat{y}$ component of $\vec{B}_{d}$ to be opposite in sign and greater in magnitude than $B_{y}$. In the hypothetical case where either $\vec{B}$ and/or $\vec{B}_{d}$ were purely along $\hat{x}$, it would not have lead to the formation of the rings, as the existence of non-zero transverse field components are crucial.

From this point on, a ring is finally formed after the red domains shrink due to the opposing $\vec{B}$ [figure \ref{fig:fig4} (b), states (3) - (5)]. The key role of an alternating $B_{y}$ field [figure \ref{fig:fig4} (b), states (0) - (1)] for the formation of 360$^{\circ}$ DW rings in a system with $m_{x}$ domains is similar to previous works in nanowires with injection pads \cite{jang2012formation}, where 360$^{\circ}$ DW were stabilized via the application of alternating external magnetic fields. Here, instead, the net magnetic field is dominated by $\vec{B}_{ext}$ or $\vec{B}_{d}$, for different steps of formation.

In this ideal simulation space, the ring eventually annhilates, in contrast with the experimental results where rings remain stable. This can be readily explained due to the presence of pinning sites, defects, and imperfections in the real sample that locally alter the magnetic energy landscape, leading to their meta-stability \cite{coey2010magnetism}.

In conclusion, we have observed 360$^{\circ}$ DW rings via XMCD-PEEM magnetic vector reconstruction, forming in a SAF exhibiting exchange bias due to IL-DMI. These textures are observed at remanence after applying IP demagnetizing field sequences where a global offset in field is added. A combination of XMCD-PEEM and XRMS show how the rings are only found when the field offset does not perfectly compensate for the IL-DMI exchange bias field present in the SAF. We propose a mechanism for the formation and stability of the rings, based on analyzing their magnetic charge distribution and topology, in combination with micromagnetic simulations. First, a non-zero net IP field parallel to the easy axis, result of external and IL-DMI fields, is key for their stability at remanence. This net field in combination with pinning sites prevents their relaxation and annihilation due to their trivial topology. Secondly, a non-zero component of the net field perpendicular to the easy axis is required for their formation. This component sets the initial direction of two 180$^{\circ}$ walls at the start of the reversal process, which subsequently combine with two other 180$^{\circ}$ walls that form afterwards. The wall component of this second set of 180$^{\circ}$ walls is opposite to the original ones, set by the strong demagnetizing field in between domains that dominate as these become closer to each other. The combination of the two sets of consecutive 180$^{\circ}$ walls results in two 360$^{\circ}$ walls of opposite chirality, leading to a 360$^{\circ}$ wall ring. A deep understanding of the effect of IL-DMI on the magnetic reversal process and domain structure of SAFs is crucial for the potential exploitation of this effect in spintronic devices.

\begin{acknowledgments}
This work was supported by UKRI through an EPSRC studentship, EP/N509668/1 and EP/R513222/1, the European Community under the Horizon 2020 Program, Contract No. 101001290 (3DNANOMAG), the MCIN with funding from European Union NextGenerationEU (PRTR-C17.I1), and the Aragon Government through the Project Q-MAD. The raw data supporting the findings of this study will be openly available at the DIGITAL.CSIC repository.

A.H.-R. acknowledges the support by Spanish MICIN under grant PID2019-104604RB/AEI/10.13039/501100011033 and by Asturias FICYT under grant AYUD/2021/51185 with the support of FEDER funds. S.R-G. acknowledges the financial support of the Alexander von Humboldt foundation. L.S. acknowledges support from the EPSRC Cambridge NanoDTC EP/L015978/1. C.D. acknowledges funding from the Max Planck Society Lise Meitner Excellence Program. The ALBA Synchrotron is funded by the Ministry of Research and Innovation of Spain, by the Generalitat de Catalunya and by European FEDER funds. We acknowledge Synchrotron SOLEIL for providing the synchrotron radiation facilities (Proposal No. 20191674). S.M. acknowledges support from EPSRC project EP/T006811/1. M.A.N and M.F. acknowledge support from MICINN project ECLIPSE (PID2021-122980OB-C54).

\end{acknowledgments}

\bibliography{bibliography}

\end{document}